\documentclass[aps,prc,twocolumn,floatfix]{revtex4}
\usepackage{epsfig,bm}
\usepackage{type1cm}
\begin{document}

\title{
Parity-Projected Shell Model Monte Carlo Level Densities 
for $fp$-shell Nuclei}
\author{C. \"Ozen$^1$, K. Langanke$^{1,2}$, G. Martinez-Pinedo$^{1}$, and D.J. Dean$^3$}
\affiliation{
$^1$
Gesellschaft f\"ur Schwerionenforschung GSI, Darmstadt, Germany \\
$^2$
Institut f\"ur Kernphysik, Technische Universit\"at Darmstadt, Darmstadt, 
Germanny \\
$^3$
Physics Division, Oak Ridge National Laboratory, Oak Ridge, TN, USA \\
}

\date{\today}


\begin{abstract}
\noindent
We calculate parity-dependent level densities for 
the even-even isotopes $^{58,62,66}$Fe and $^{58}$Ni and the odd-$A$
nuclei $^{59}$Ni and $^{65}$Fe 
using the
Shell Model Monte Carlo method. We perform these calculations 
in the complete $fp$-$gds$ shell-model space using a
pairing+quadrupole residual interaction.
We find that, due to pairing of identical nucleons, the  low-energy spectrum is dominated by positive parity states. 
Although these pairs break at around the same excitation energy
in all nuclei, the energy dependence of the ratio of negative-to-positive 
parity level densities depends strongly on the particular nucleus of interest.
We find equilibration of both parities 
at noticeably lower excitation energies 
for the odd-$A$ nuclei $^{59}$Ni and $^{65}$Fe than for the neighboring 
even-even nuclei $^{58}$Ni and $^{66}$Fe.
\end{abstract}

\maketitle

\section{\label{sec:intro}Introduction and Motivation}

Nuclear level densities play an important role in theoretical estimates
of nuclear reaction rates needed in various applications 
including astrophysical nucleosynthesis
processes like the s-, r-, and rp-process \cite{Thielemann} They
also contribute
 to one of the largest uncertainties in the 
cross section determinations used for
large-scale nucleosynthesis networks 
\cite{Thielemann,Thielemann05,Rauscher06,Rauscher}. 
Typically, nuclear level densities are described in these
astrophysical studies by using the backshifted Fermi gas 
model of Gilbert and Cameron \cite{Gilbert}. 
This model extends the non-interacting Fermi gas model of Bethe
\cite{Bethe}
by considering pairing among like nucleons via a backshift of the
excitation energy $E$. This backshift accounts for 
the energy required to break nucleon pairs. Furthermore, the 
astrophysical applications
assume an empirical ansatz for the angular momentum distribution
of the levels and consider equilibration of both parities at the
energies of interest \cite{Thielemann,Rauscher}. 

For astrophysical applications, one is often 
interested in the nuclear level density at
rather low excitation energies. For example, the typical 
neutron energies in r-process nucleosynthesis reach only up to 
a few MeV \cite{Thielemann,Rauscher}.
At such energies, nuclear structure and pairing 
effects strongly influence the level density 
and an equilibration of both parities
is quite unlikely. In particular, in even-even nuclei, pairing
among identical nucleons generates only positive parity states even
when single-particle states of opposite parity are present. 
Therefore, negative parity states should be 
suppressed in even-even nuclei
at low energies due to both pairing and the underlying 
single-particle structure of the mean field
which groups states of the same parity, at least for nuclei with
mass numbers $A$ smaller than about 100. The parity equilibration
is governed by the energy scales associated with 
pair-breaking and the shell gap between opposite-parity states
near the Fermi surface.  While the latter is typically of 
order 5-6 MeV for intermediate mass
nuclei, the former strongly depends on the nuclear structure. 

We investigate in this paper the competition
of pairing and single-particle structure on the parity dependence
of the level density. We perform Shell Model Monte Carlo
calculations of the even-even nuclei $^{58,62,66}$Fe and $^{58}$Ni
and the odd-$A$ nuclei $^{59}$Ni and $^{65}$Fe in the complete
$fp$-$gds$ model space, where the single-particle states in the $fp$
shell have negative parity, while those of the $gds$ shell have 
positive parity. As we shall see, residual interactions
among the protons and neutrons also influence the level density. 

\section{\label{sec:form}Formalism}

The Shell Model Monte Carlo (SMMC) approach allows the determination of
nuclear properties at finite temperature in unprecedentedly large model spaces,
considering the important correlations among the valence nucleons
\cite{Johnson,report,dean95}.
The SMMC method
describes the nucleus by a canonical ensemble at temperature  
$T=\beta^{-1}$ and employs a Hubbard-Stratonovich linearization 
\cite{Hubbard} of the  
imaginary-time many-body propagator, $e^{-\beta H}$, to express  
observables as path integrals of one-body propagators in fluctuating  
auxiliary fields \cite{Johnson,Lang}. Since Monte Carlo techniques  
avoid an explicit enumeration of the many-body states, they can be  
used in model spaces far larger than those accessible to conventional  
methods. The notorious sign problem that plagues Monte Carlo studies
of Fermionic systems can be avoided within the SMMC by adopting
a pairing+quadrupole force as residual interaction \cite{report}. 

There are already several successful studies of level densities
using the SMMC approach. Ormand \cite{Ormand}, as well as
Nakada and Alhassid \cite{Nakada97}, calculated the level density
for selected even-even nuclei, while Ref. \cite{Langanke98}
extended this work to odd-A and odd-odd nuclei. 
Using projection techniques, Alhassid and collaborators succeeded
in studying the parity dependence \cite{Nakada97} and the angular-momentum
dependence \cite{Alhassid06} of the level density for intermediate-mass nuclei.
Recently, Ref. \cite{Langanke06} explored the influence of pairing correlations
and their energy dependence on nuclear level densities.
All these approaches are based on the ability of the SMMC to calculate
expectation values of an observable at temperature $T$ as the thermal
average of a canonical ensemble (with fixed numbers of protons and neutrons).
Thus for the energy excitation function, one has
\begin{equation}
E(\beta)   = 
\frac{{\rm Tr} \left[ H e^{-\beta H} \right]}{{\rm Tr} \left[ e^{-\beta H} \right]}=
\frac{{\rm Tr} \left[ H e^{-\beta H} \right]}{Z(\beta)}\;,
\end{equation}
where $Z$ is the partition function, $\beta$ the inverse temperature,
and $H$ the nuclear Hamiltonian.
Using the parity projection 
operators $P_\pm = (1 \pm P)/2$, where $P$ is the 
parity operator, one is able to calculate the ratio of the parity-projected
partition function to the total partition function \cite{Nakada97}

\begin{equation}
Y_\pm (\beta) = \frac{Z_\pm (\beta)}{Z (\beta) } = \frac{{\rm Tr}[P_\pm e^{-\beta H}]}{{\rm Tr}[e^{-\beta H}]}
\end{equation}
from which one can then extract the parity-projected energy excitation functions

\begin{eqnarray}
E_\pm(\beta)  & = &
-\frac{d {\rm ln} \, Y_\pm (\beta) }{d \beta} + E (\beta) \nonumber \\ & = &
\frac{{\rm Tr} \left[ H P_\pm e^{-\beta H} \right]}{Z_\pm(\beta)} \nonumber \\ & = &
\frac{\int dE' e^{-\beta E'} E \rho_\pm(E')}{Z_\pm(\beta)}\;,
\label{eq:parity}
\end{eqnarray}
where we have introduced the parity-projected partition functions
$Z_\pm$ and level densities $\rho_\pm$.

To obtain $\rho_\pm$ from Eq. \ref{eq:parity} requires an inverse
Laplace transform which we treat within the saddle-point approximation:
\begin{equation}
\rho_\pm(E) = \frac{e^{\beta E_\pm + ln Z_\pm(\beta)}}{\sqrt{-2\pi
\frac{dE_\pm(\beta)}{d\beta}}}\;,
\end{equation}
where $\beta=\beta(E_\pm)$ is obtained by inverting $E_\pm(\beta)$.
An analog relation holds between the total level density $\rho(E)$
and the energy excitation function $E(\beta)$.

\section{\label{sec:res}Results}

Our SMMC calculations were performed in the complete $fp$-$gds$
model space with 50 valence orbitals for both protons and neutrons.
The single-particle energies and the residual pairing+quadrupole interaction
were adopted from previous SMMC studies that successfully
explored the competition of isovector pairing versus
quadrupole deformation in the $A \sim 80$ mass region \cite{Langanke03},
and as a function of temperature in selected nuclei. These previous
calculations clearly identified the breaking of pairs of 
identical nucleons in even-even nuclei
around $T \approx 0.7$ MeV \cite{Langanke05}.

\begin{figure}[htb]
\begin{center}
  \leavevmode
    \rotatebox{0}{\scalebox{0.9}{\includegraphics{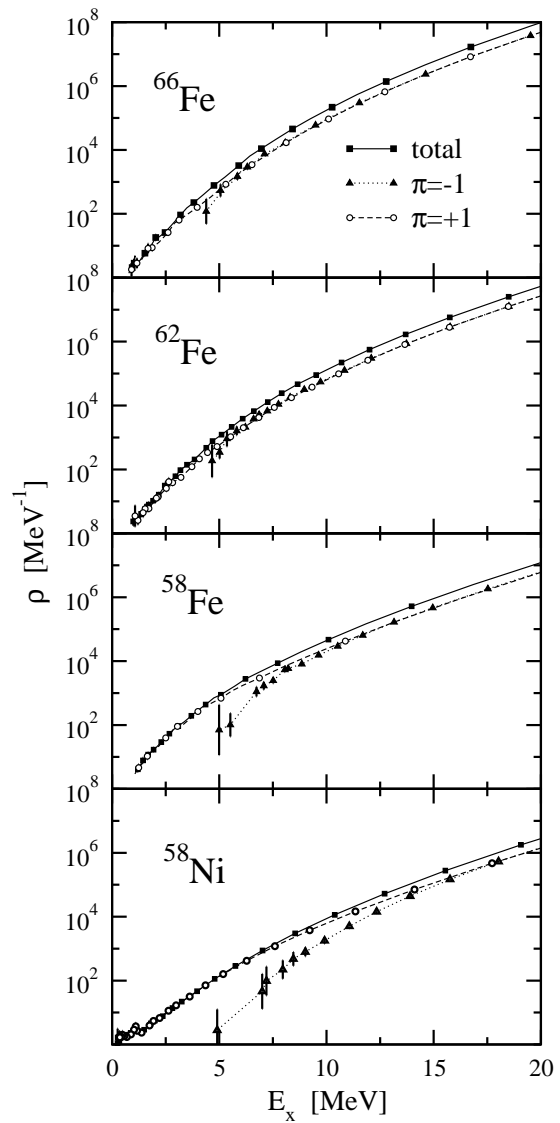}}}
    \caption{Parity-projected level densities for
               $^{58,62,66}$Fe and $^{58}$Ni.
      }
    \label{fig:rho-even}
  \end{center}
\end{figure}

\begin{figure}[htb]
\begin{center}
  \leavevmode
    \rotatebox{0}{\scalebox{0.9}{\includegraphics{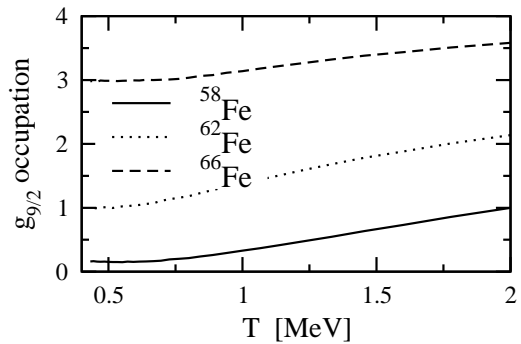}}}
    \caption{Average SMMC occupation numbers of the $g_{9/2}$ neutron orbitals 
      for $^{58,62,66}$Fe
      as function of temperature.}
    \label{fig:g92}
  \end{center}
\end{figure}

In the following, we discuss various results of these calculations. 
Figure \ref{fig:rho-even}  shows the total and
parity-projected SMMC level densities for
$^{58,62,66}$Fe and $^{58}$Ni. We observe that, at modest excitation energies, 
the total level density increases with the mass number. This is expected
from a Fermi gas approximation to the level density, which scales
like $\rho \sim {\rm exp} \{ 2 \sqrt{a E} \}$ with the level density parameter
$a$ proportional to the mass number $A$ \cite{Thielemann}.
Interestingly, we observe
an equilibration of both parities at $E \approx$ 11-12
in $^{58}$Fe, while equilibration occurs already around 
$E \approx 6$ MeV in $^{62,66}$Fe.
This observation simply reflects the fact that it is energetically
more costly to promote neutrons to the $gds$ shell for $^{58}$Fe,
which requires neutrons to bridge the gap from the $p_{3/2}$ to the $g_{9/2}$
orbital, than for $^{62,66}$Fe.
Nevertheless, it is not simply the  occupation number of the
$g_{9/2}$ orbital which matters. This quantity is shown in Fig.
\ref{fig:g92}. For $^{58}$Fe, the SMMC predicts rather little average
occupation of the $g_{9/2}$ orbital,
which increases from 0.2 neutrons in the ground state to 0.3 at $T=1$ MeV
(corresponding to an excitation energy of about $E \sim 6$ MeV) to 1.0 at $T=2$ MeV
($E \sim 23$ MeV). 
On the other hand, one finds an occupation number
of order 1.0 in the $^{62}$Fe ground state, which grows to 2.1 at $T=2$ MeV.
For $^{66}$Fe, the residual interaction promotes neutrons 
to the $g_{9/2}$ orbital, yielding a $g_{9/2}$ occupation number
of about 3.0. The occupation grows only mildly with temperature and reaches
about 3.6 at $T=2$ MeV.
We note that the residual interaction mainly scatters (neutron) pairs from the
$fp$ shell to the $gds$ shell. Although these correlations are important
for the nuclear structure, they do not change the parity. This is clearly
seen in a comparison between $^{62}$Fe and $^{66}$Fe, which shows quite
a similar ratio of negative-to-positive level densities (Fig. \ref{fig:rho-even}),
 while clearly these nuclei have distinct occupation 
numbers of the $g_{9/2}$ orbital (Fig. \ref{fig:g92}). While a promotion
of nucleons (neutrons) is required to make negative-parity states
in our model space, the strong pairing
correlations also have to be overcome. 

As in \cite{Langanke96,Langanke05}, we define the pairing strength
as the sum over all matrix elements of the pair matrix
\begin{equation}
M_{\alpha,\alpha'}=
\langle A^\dagger (j_a,j_b) A (j_c,j_d) \rangle,
\label{eq:pair}
\end{equation}
with the $J=0$ pair operator
\begin{equation}
A^\dagger (j_a,j_b) =
\frac{1}{\sqrt{1+\delta_{ab}}}
\left[a^\dagger_{j_a}
\times a^\dagger_{j_b}\right]^{JM=00}\;,
\end{equation}
where $a^\dagger_{j_a}$ creates a nucleon in the orbital $a$ with angular
momentum $j_a$ 
\footnote{There is a misprint in \cite{Langanke05,Langanke06}.
As in the present work, the plotted pairing strengths are 
the sum over the matrix elements
of the pair matrix $M$, not the sum over its eigenvalues.}. 
Since only genuine
pair correlations are of interest, we subtract the pure `mean-field'
values, i.e. those obtained without residual interaction. 

\begin{figure}[htb]
\begin{center}
  \leavevmode
    \rotatebox{0}{\scalebox{0.9}{\includegraphics{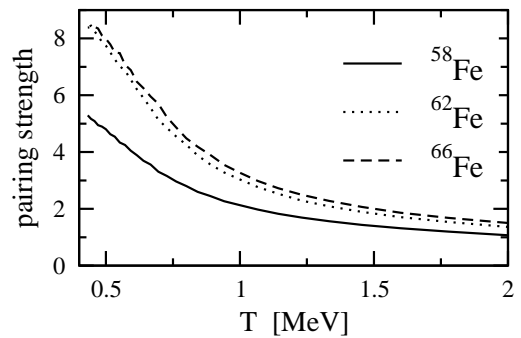}}}
    \caption{Pairing strength for neutrons in $^{58,62,66}$Fe
      as function of temperature. As explained in the text, the pairing
      strength is defined as the sum of the matrix elements of the 
     pair 
      matrix \ref{eq:pair} with the `mean-field' value subtracted.}
    \label{fig:pairs}
  \end{center}
\end{figure}

As is shown in Fig. \ref{fig:pairs},
the pairing strengths 
decrease strongly with temperature, and are reduced to about half of
the ground-state values 
around $T=0.8$ MeV (corresponding to $E \approx 4$ MeV). 
This breaking of pairs is accompanied by a peak in the specific heat
(e.g. \cite{Langanke05}). We note that the proton pairing strengths
are quite similar for all three iron isotopes (e.g. \cite{Langanke98}).

For $^{62}$Fe and $^{66}$Fe the ratio of negative-to-positive level densities
is quite similar and equilibration is reached at excitation energies 
around 6 MeV; i.e., once the pairs with positive parities are sufficiently
broken. This apparently requires more energy than the gap between 
the last occupied neutron orbital and the $g_{9/2}$ orbital
with its large degeneracy. Hence, once
pairs are broken, many negative-parity states can be formed in $^{62}$Fe and
$^{66}$Fe. The situation is obviously different for $^{58}$Fe where
the gaps from the last occupied proton ($f_{7/2}$) and neutron ($p_{3/2}$)
orbitals to the $g_{9/2}$ orbital are larger than 6 MeV. Thus, even 
at the energy at which 
pairs are broken, it is more likely to form positive-parity states with unpaired
nucleons in the $fp$ shell than negative-parity cross-shell states.

Ref.~\cite{Bertsch} describes a model to estimate the ratio
of negative-to-positive level densities on the basis of the independent particle
model and a BCS treatment of pairing. It appears that this model slightly 
underestimates this ratio. This is  due to the fact that 
the residual interaction mixes orbitals from adjacent 
shells with different parities at lower energies
than obtained in the independent particle model.

Although they have the same neutron number, the isotones 
$^{58}$Fe and $^{58}$Ni differ in their
single-particle structure, and this shows in their level densities.
The fact that $^{58}$Fe has four neutrons in the $p_{3/2}$ orbital,
compared to two in the case of $^{58}$Ni, allows for a larger amount of
negative-parity states at low energies in $^{58}$Fe, since these states are generated by 
configurations with one and three neutrons in the $gds$-orbitals for 
$^{58}$Fe but with only one neutron for $^{58}$Ni.
As a consequence, the parity unbalance
in the level density persists to higher excitation energies in $^{58}$Ni
(about $E_x = 16$ MeV, see Fig. \ref{fig:rho-even}) than in $^{58}$Fe. 
If the total level densities
of the two nuclei are compared, the one for $^{58}$Fe is noticeably larger
at a given excitation energy than the one for $^{58}$Ni. At modest 
excitation energies, this difference can be approximately accounted for
by a constant energy shift of about 3 MeV, simulating the shell gap
in $^{58}$Ni.

\begin{figure}[htb]
\begin{center}
  \leavevmode
    \rotatebox{0}{\scalebox{0.9}{\includegraphics{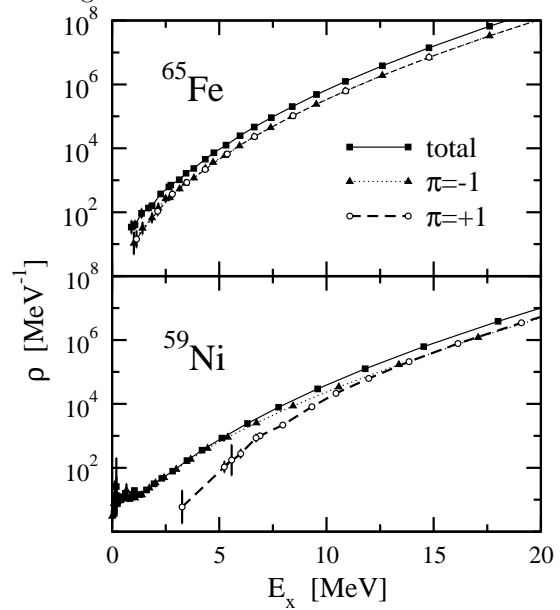}}}  
    \caption{Parity-projected level densities for
              $^{65}$Fe and $^{59}$Ni.
      }
    \label{fig:rho-odd}
  \end{center}
\end{figure}

While the structure of even-even nuclei is
strongly dominated by pairing at low excitation energies, thus 
leading to the dominance of positive-parity
over negative-parity states, the situation is different for odd-$A$ nuclei.
In this case, the unpaired nucleon is not hindered by pairing 
and can, depending on the single-particle structure, 
occupy negative-parity states ($fp$ shell)
or positive-parity states ($gds$ shell). To investigate
the differences  between even-even and odd-$A$ nuclei,
we also calculated the SMMC level densities for $^{59}$Ni and
$^{65}$Fe. For $^{59}$Ni, the unpaired neutron occupies a $p_{3/2}$ orbital.
Despite the fact that this neutron is not hindered by pairing, it will
nevertheless, at low excitations energy, mainly 
occupy orbitals in the $fp$ shell; consequently, negative-parity 
states dominate the $^{59}$Ni level
density at low excitation energies. This is indeed born out of our
SMMC calculation (Fig. \ref{fig:rho-odd}), which yields a balance between
negative- and positive-parity 
level densities at excitation energies around 13 MeV. Comparing with 
$^{58}$Ni (Fig.~\ref{fig:rho-even}), where the 
equilibration energy is about 16 MeV, we see that
neutron pairing, e.g., the residual interaction,  indeed has 
a signifant effect in the equilibration of positive and negative
parities. 

The neutron single-particle
structure of $^{65}$Fe corresponds to a single hole in 
the $fp$ shell.
Thus, low-energy excitations can be achieved 
by either changing the hole structure
in the $fp$ shell or by promoting the neutron to the $g_{9/2}$ orbital,
which, however, has the opposite (positive) parity to the ground state.
As a consequence, one expects equal distribution of 
negative- and positive-parity level densities already 
at low excitation energies. This is indeed observed
in our SMMC calculations (Fig. \ref{fig:rho-odd}), 
where we find the same level densities
for both parities down to $E_x \approx 1.5$ MeV, corresponding
to the lowest temperature for which we have been able to perform
SMMC calculations in the odd-A systems. (An odd-A sign problem 
occurs even in the presence of good sign interactions \cite{report}.)

\section{\label{sec:conc}Conclusions}

In summary, we calculated parity-projected level densities for several
even-even and odd-$A$ $fp$-shell nuclei using the Shell Model Monte Carlo
approach. For even-even nuclei, we confirm that the low-energy spectrum
is dominated by pairing among nucleons, resulting in a dominance
of the positive-parity level density. Although the pairs break
at excitation energies of a few MeV, the contribution of negative-parity
states depends strongly on the single-particle structure of the
nuclei. If the Fermi energies of protons and neutrons are relatively
well separated from orbitals with the opposite parity, an equal 
distribution between parities in the level densities is 
achieved at higher excitation energies than for nuclei with Fermi 
energies close to the oscillator shell closure.  As examples for 
this difference, we presented parity-projected level
densities for $^{58}$Fe, $^{58}$Ni, and $^{66}$Fe. The same general trend can
be observed for odd-$A$ nuclei ($^{59}$Ni and $^{65}$Fe). However,
as these nuclei have an unpaired neutron, the balance between negative- 
and positive-parity level densities is achieved at 
somewhat lower excitation
energies than in the neighboring even-even nuclei.

The SMMC approach has again been proven to be a powerful tool to microscopically
study nuclear level densities. In the future, it will be used to explore
the angular-momentum dependence of the level density. First steps
towards this goal have been presented in \cite{Alhassid06,vanHoucke}.

\acknowledgments
Oak Ridge National Laboratory is managed by UT-Battelle, LLC, for the 
U.S. Department of Energy under contract DE-AC05-00OR22725. 
Computational resources were provided by the National Center
for Computational Sciences at ORNL.

\end{document}